%% file: main.tex
\documentclass[sigconf,screen]{acmart} 
\acmConference[ICSE 2024]{46th International Conference on Software Engineering}{April 2024}{Lisbon, Portugal}
\usepackage[normalem]{ulem}
\usepackage[utf8]{inputenc}
\usepackage{amsmath}  
\usepackage{graphicx}
\usepackage{comment}
\usepackage{multirow}
\usepackage{enumitem}
\usepackage{array}
\usepackage{pifont}
\usepackage{xcolor}
\usepackage{balance}
\usepackage{mathtools}
\usepackage{xspace}
\usepackage{url}
\usepackage{eucal}
\usepackage{amsfonts}
\usepackage{color}
\usepackage{booktabs}
\usepackage{bbding}
\usepackage{float}

\usepackage{arydshln}

\usepackage{graphicx}
\usepackage{hyperref}
\usepackage{url}
\usepackage{multirow}
\usepackage{comment}
\usepackage{colortbl}
\usepackage{graphicx}
\usepackage{subcaption}
\usepackage[linewidth=0.1pt]{mdframed}
\usepackage[linesnumbered,ruled,lined]{algorithm2e}

\usepackage{wrapfig,lipsum,booktabs}
\usepackage{subfloat}
\usepackage[export]{adjustbox}

\AtBeginDocument{%
  \providecommand\BibTeX{{%
    \normalfont B\kern-0.5em{\scshape i\kern-0.25em b}\kern-0.8em\TeX}}}

\newcommand{\revised}[1]{\textcolor{black}{#1}}
\newcommand{\revisedmar}[1]{\textcolor{black}{#1}}

\pdfpagewidth=\paperwidth
\pdfpageheight=\paperheight

\makeatletter
\newcommand{\showfontsize}{\f@size{} pt}
\newcommand\usemm[1]{%
  \strip@pt\dimexpr0.3514598\dimexpr #1\relax\relax mm%
}
\newcommand\usein[1]{%
  \strip@pt\dimexpr0.013837\dimexpr #1\relax\relax in%
}
\makeatother

\usepackage{graphicx}
\usepackage{hyperref}
\usepackage{url}
\usepackage{multirow}
\usepackage{comment}
\usepackage{colortbl}
\usepackage{graphicx}
\usepackage{tcolorbox}
\usepackage{arydshln}
\usepackage{algpseudocode}  
\usepackage{amsmath}  
\usepackage{caption} 
\usepackage{algpseudocode}
\usepackage{float}
\usepackage{soul}

\usepackage{xcolor}
\usepackage{enumitem}
\usepackage{amsmath}

\definecolor{dark-red}{RGB}{255,0,0}
\definecolor{dark-green}{RGB}{0,200,0}

\newboolean{showcomments}
\setboolean{showcomments}{true}
\ifthenelse{\boolean{showcomments}}

\setcopyright{none}

\acmDOI{}
\copyrightyear{2024} 
\acmYear{2024} 
\setcopyright{rightsretained} 
\acmConference[ICSE '24]{2024 IEEE/ACM 46th International Conference on Software Engineering}{April 14--20, 2024}{Lisbon, Portugal}
\acmBooktitle{2024 IEEE/ACM 46th International Conference on Software Engineering (ICSE '24), April 14--20, 2024, Lisbon, Portugal}\acmDOI{10.1145/3597503.3639222}
\acmISBN{979-8-4007-0217-4/24/04}
\begin{document}

\pagestyle{fancy}
\fancypagestyle{preContent}{
    \renewcommand\headrulewidth{0pt}
    \fancyfoot[C]{\thepage}
}
\pagestyle{preContent}
\pagenumbering{arabic}

\settopmatter{printacmref=false}

\title{\revisedmar{Multi-LLM Collaboration + Data-Centric Innovation =  \\2x Better Vulnerability Repair}}

\author{Xin Zhou}
\affiliation{%
  \institution{Singapore Management University}
  \country{Singapore}}
\email{xinzhou.2020@phdcs.smu.edu.sg}

\author{Kisub Kim}
\affiliation{%
  \institution{\revised{Singapore Management University}}
  \country{\revised{Singapore}}}
  \authornote{Corresponding author.}
\email{kisubkim@smu.edu.sg}

\author{Bowen Xu}
\affiliation{%
  \institution{North Carolina State University}
  \country{USA}}
\email{bxu22@ncsu.edu}

\author{DongGyun Han}
\affiliation{%
  \institution{Royal Holloway, University of London}
  \country{United Kingdom}}
\email{donggyun.han@rhul.ac.uk}

\author{David Lo}
\affiliation{%
  \institution{Singapore Management University}
  \country{Singapore}}
\email{davidlo@smu.edu.sg}

\input{Sections/0_abstract}

\maketitle

\input{Sections/1_introduction}

\input{Sections/2_background}
\input{Sections/3_approach}

\input{Sections/4_setup}

\input{Sections/5_result}

\input{Sections/6_discussion}

\input{Sections/7_related}

\input{Sections/8_conclusion}

\balance
\bibliographystyle{ACM-Reference-Format}
\bibliography{sample-base}

\end{document}

%% file: Sections/0_abstract.tex
\begin{abstract}

The advances of deep learning (DL) have paved the way for automatic software vulnerability repair approaches, which effectively learn the mapping from the vulnerable code to the fixed code. Nevertheless, existing DL-based vulnerability repair methods face notable limitations: 1) they struggle to handle lengthy vulnerable code, 2) they treat code as natural language texts, neglecting its inherent structure, and 3) they do not tap into the valuable expert knowledge present in the expert system.

To address this, we propose VulMaster, a Transformer-based neural network model that excels at generating vulnerability repairs through data-centric innovation. Specifically, VulMaster introduces the utilization and combination of various types of input data, including complete vulnerable code of any size, vulnerable code structures, and expert knowledge from the CWE system.
Additionally, VulMaster leverages the collaboration between two Large Language Models (LLMs), CodeT5 and ChatGPT: CodeT5 acts as the customizable backbone LLM, fine-tuned with the training data, while ChatGPT supplements by providing missing relevant inputs to CodeT5.
We evaluated VulMaster on a real-world C/C++ vulnerability repair dataset comprising 1,754 projects with 5,800 vulnerable functions. 
The experimental results demonstrated that VulMaster exhibits substantial improvements compared to the learning-based state-of-the-art vulnerability repair approach. 
Specifically, VulMaster improves the EM, BLEU, and CodeBLEU scores from 10.2\% to 20.0\%, 21.3\% to 29.3\%, and 32.5\% to 40.9\%, respectively.

\end{abstract}

%% file: Sections/1_introduction.tex
\section{Introduction} 
\label{section:introduction}

As the software landscape expands rapidly~\cite{github_report}, the number of software vulnerabilities has increased correspondingly~\cite{homaei2017seven}. 
According to Common Vulnerabilities and Exposures~\cite{cve_home}, the yearly discovery of vulnerabilities in 2022 sets a new record with 26,448 software vulnerabilities reported, marking a notable 59\% increase compared to 2021~\cite{security_report2}.
Recently, many automatic vulnerability prediction approaches~\cite{zhou2019devign,chakraborty2021deep,nguyen2019deep} are proposed to identify the vulnerability. However, when a software vulnerability is detected, it becomes of utmost importance to promptly rectify and resolve it in order to minimize the potential risks of exploitation. However, addressing software vulnerabilities often requires specialized expertise, and the existing pool of experienced developers is insufficient to tackle the extensive number of vulnerabilities found in millions of software systems~\cite{ji2018coming}. 
Furthermore, the manual resolution of vulnerabilities is a time-consuming process; the GitHub 2020 security report finds that it takes 4.4 weeks to fix a vulnerability after its identification~\cite{forsgren20212020}. As a result, there is an urgent demand for automated methods to repair vulnerabilities.

A number of program analysis-based vulnerability repair approaches~\cite{DBLP:conf/ccs/MaLLD16,senx,saver,memfix,van2018static,zhang2022program,gao2019crash,cpr,extractfix} have been proposed. Although effective, program analysis-based approaches are often tailored for specific vulnerabilities or not applicable to certain types. In contrast, learning-based vulnerability repair approaches are not constrained to fixing specific vulnerability types. In this study, we aim to advance learning-based vulnerability repair approaches.
Recently, several learning-based automatic vulnerability repair (AVR) approaches that take a vulnerable function with its CWE type (e.g., CWE-119) as input and generate the repaired function as as output have been proposed~\cite{vrepair,vulrepair}.
For instance, VRepair~\cite{vrepair} is pre-trained on a large bug-fixing corpus and then trained to repair vulnerabilities. 
On the other hand, VulRepair directly utilizes a pre-trained model, named CodeT5~\cite{CodeT5}.
By harnessing this off-the-shelf pre-trained model, VulRepair achieves state-of-the-art performance in learning-based vulnerability repair~\cite{vulrepair}. 
Despite the promising outcomes, VRepair and VulRepair encounter three major \textbf{challenges}.

\begin{figure*}[t] 
\centering 
\includegraphics[width=1\linewidth]{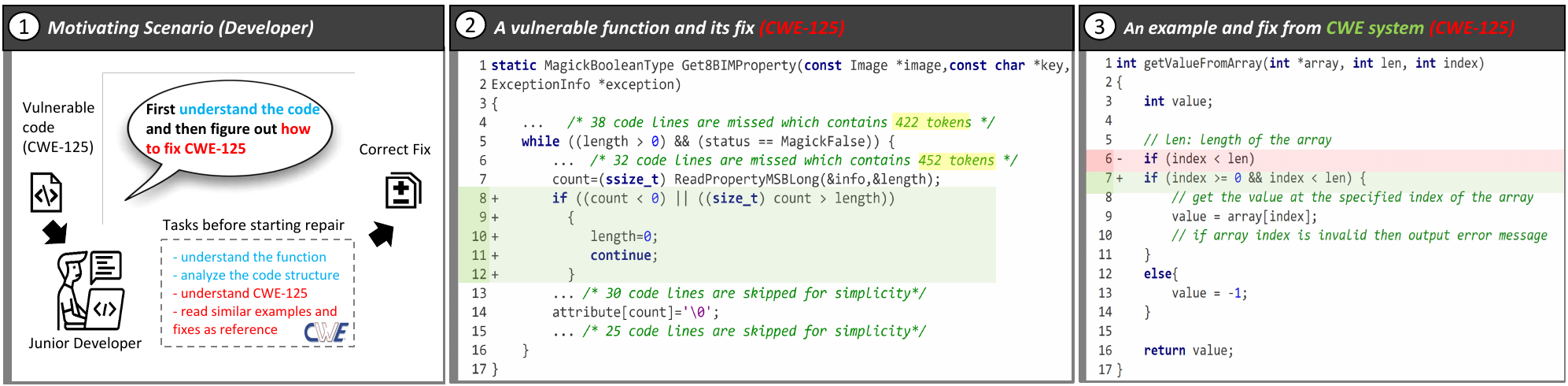} 
\caption{A motivating example. \ding{182} the process of how junior developers repair vulnerability; \ding{183} a vulnerable function and its fixes from the {\tt ImageMagick} project;  \ding{184} a vulnerable example and its fixes from the CWE website.}
\label{fig:motivation} 
\end{figure*}

\textbf{Understanding the entire vulnerable code:}
VRepair and VulRepair rely on the Transformer model, and its computational cost significantly increases with longer inputs. The attention matrix computations (i.e., the primary calculations) within the Transformer model scale quadratically with the length of the input sequence~\cite{10.1145/3530811}. As a result, Transformer-based models inevitably restrict the length of input code snippets, leading to truncation of longer parts. For example, VulRepair imposes a maximum limit of 512 code tokens. 
However, real-world vulnerable codes often exceed this limit: 44.9\% of the vulnerable code in VulRepair's evaluation dataset~\cite{vulrepair} surpasses 512 tokens. 
Failing to provide and understand the entire vulnerable code to the model reduces the likelihood of accurately addressing the vulnerability.

\textbf{Understanding the structures of the vulnerable code:}
Both VRepair and VulRepair treat code in a similar manner to natural language texts, disregarding the incorporation of structural information. 
Natural language possesses a loosely structured nature, enabling words to be arranged in different orders while maintaining grammatical correctness~\cite{mirault2018you,nl_loose}.
In contrast, programming languages exhibit a higher level of structure. 
Neglecting code structures could reduce the probability of effectively resolving the vulnerability in both VRepair and VulRepair. 
Prior research has demonstrated the benefits of including code structures, such as the Abstract Syntax Tree (AST), in tasks like code clone detection~\cite{zhang2023efficient,meng2020deep}, code search~\cite{niu2022spt}, and bug comprehension~\cite{mahbub2023explaining}. Following these prior studies, we also include the AST as a part of the model’s input to improve its grasp of the structural characteristics of code.

\textbf{Leveraging the expert knowledge:}
Software vulnerabilities do not exist in isolation. 
The Common Weakness Enumeration (CWE) system~\cite{cwe} offers a comprehensive catalog of common software weakness types.
By utilizing the CWE system, one may gain access to accurate descriptions, vulnerable code examples, and information on closely related vulnerabilities pertaining to specific vulnerability types. 
In order to tap into the expert knowledge provided by the CWE system, both VRepair~\cite{vrepair} and VulRepair~\cite{vulrepair} incorporate CWE types into their input data. The CWE system, however, offers a wealth of expert knowledge that extends beyond CWE types. This includes CWE names and vulnerable code examples shared on the web pages. Our approach aims to better utilize this abundant knowledge within the CWE systems.

To tackle these challenges, we propose \textbf{VulMaster}, a novel automatic vulnerability repair approach based on data-centric innovation. Specifically, VulMaster introduces the utilization and combination of various types of input data, including complete vulnerable code of any size, vulnerable code structures, and expert knowledge from the CWE system.
To capture the structural aspects of the vulnerable code, VulMaster utilizes the AST as part of its input.
It extensively utilizes expert knowledge of the CWE system, including the vulnerability type name, typical vulnerable code examples, and extra information on other closely related vulnerabilities.
Technically, VulMaster follows the idea of the \textit{Fusion-in-Decoder (FiD)} framework~\cite{FID} to overcome the limitations associated with the input length of Transformer-based models, enabling it to effectively process and fuse diverse data inputs.

In addition to fusing diverse inputs, VulMaster leverages the collaboration between two Large Language Models (LLMs), CodeT5~\cite{CodeT5} and ChatGPT~\cite{chatgpt} to enhance its effectiveness: CodeT5 acts as the customizable backbone LLM of VulMaster, fine-tuned with training data, while ChatGPT supplements by providing relevant inputs that may initially be absent for CodeT5. Specifically, ChatGPT is employed to generate fixes for typical vulnerable code examples sourced from CWE web pages, where fixes may not exist initially. These generated fixes are subsequently integrated into the input data for CodeT5.
We believe this collaboration harnesses the strengths of these two distinct LLMs: CodeT5 can be heavily customized by the training data through fine-tuning, while ChatGPT excels at generating code data when the context is clear.

\begin{figure*}[t] 
\centering 
\includegraphics[width=0.9\linewidth]{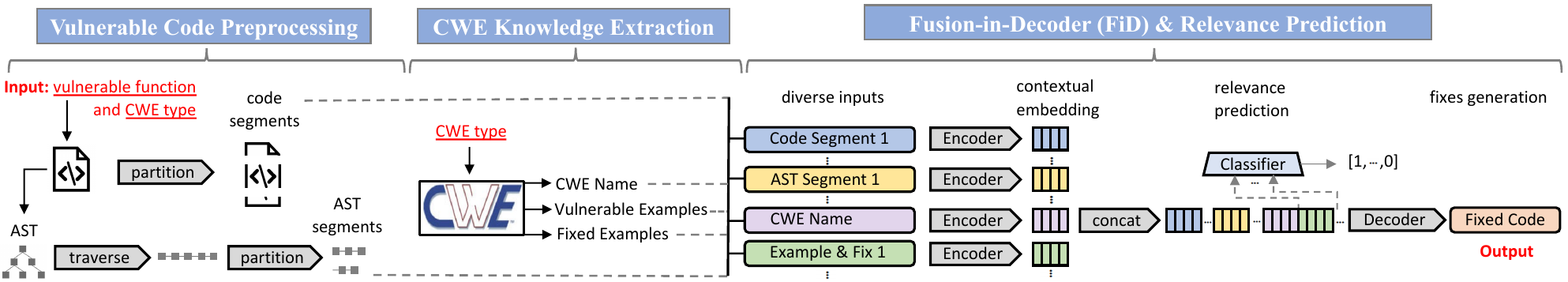} 
\vspace{-0.3cm}
\caption{Overall Framework of VulMaster.}
\vspace{-0.4cm}
\label{fig:framework} 
\end{figure*}

We evaluate VulMaster on a real-world C/C++ vulnerability dataset used in previous studies~\cite{vrepair,vulrepair,cvefixes,fan2020ac}, which consists of 5,800 function-level unique vulnerability fixes collected from 1,754 large-scale open-source software projects.
The experimental results illustrate that VulMaster enhances the EM, BLEU, and CodeBLEU scores from 10.2\% to 20.0\%, 21.3\% to 29.3\%, and 32.5\% to 40.9\%, respectively, compared to the state-of-the-art learning-based vulnerability repair solution.
In summary, our contributions are as follows:
\begin{itemize}[leftmargin=*]

\item [$\bullet$] To the best of our knowledge, we are the first to introduce the FiD framework for the field of software engineering. More importantly, we initiate how it should be leveraged to effectively mitigate the input length limitations of Transformer-based models and integrated with other valuable aiding information.

\item [$\bullet$] 
As far as we know, we are among the first to introduce collaborations between different Large Language Models (LLMs), particularly when each LLM is utilized according to its strengths and characteristics—such as fine-tuning for CodeT5 and (zero-shot) prompting for ChatGPT. The advantages of both LLMs are combined in the collaboration: CodeT5 can be heavily customized to effectively absorb the knowledge in a large amount of training data, while ChatGPT can generate the missing data for CodeT5 when the context is clear.

\item [$\bullet$] We discover more diverse and valuable information such as vulnerable code structures and expert knowledge for CWEs, which can boost the effectiveness of repairing vulnerability.

\item [$\bullet$] We further identify and address a hidden label leakage issue in the datasets used in the prior work, which can contribute to a more precise evaluation of future studies.

\end{itemize}

%% file: Sections/2_background.tex
\section{Preliminaries and Motivation}
\label{sec:background}

\subsection{Motivating Example}

In \ding{182} of Figure~\ref{fig:motivation}, we show an example to analyze how junior security engineers repair a vulnerable function of CWE-125 (e.g., the vulnerable code before fixing shown in \ding{183} of Figure~\ref{fig:motivation}) and explain our two motivations below. 

\vspace{0.05cm}
\noindent
\textbf{(1) Understanding the vulnerable function is the first thing.} 
Instead of directly crafting an appropriate fix, the security developer may first undertake a meticulous examination of the vulnerable function
to acquire a comprehensive understanding of the code and its weaknesses. 
While human developers can manage to grasp lengthy vulnerable functions, it poses a challenge for automatic tools based on Transformer models. For instance, as depicted in \ding{183} of Figure~\ref{fig:motivation}, the vulnerable function from~\cite{example_commit} consists of over 800 code tokens before the vulnerable code lines, surpassing the input limit (512 tokens) of the learning-based SOTA VulRepair. 
Consequently, VulRepair failed to generate an accurate fix.

\vspace{0.1cm}
\noindent
\textbf{(2) CWE knowledge is helpful in inspiring the repair.}
To enhance the understanding of software vulnerabilities, the junior security developer may refer to the name/description and typical vulnerable examples of CWE-125 on the CWE website.
From the example shown in \ding{184}, one typical cause of CWE-125 is forgetting to ensure the array index is not a negative number. 
Upon identifying this pattern, the developer can revisit the target vulnerable function in \ding{183}: \textit{Line 7} of the target function obtains a value for the variable {\tt count}, which is later used as an index in \textit{Line 14}. However, the function does not check whether {\tt count} is a negative value. 
As a result, a part of the ground-truth fix (in \ding{183} of Figure~\ref{fig:motivation}) could be inspired by the typical vulnerable example provided by the CWE website, i.e., checking whether the {\tt count} is negative.

\subsection{Background}

\noindent\textbf{Task Definition.}
In line with previous research~\cite{vrepair,vulrepair}, learning-based automatic vulnerability repair (AVR) is formulated as a sequence-to-sequence problem: $(X_i, T_i) \to Y_i$. 
Specifically, given a vulnerable code snippet $X_i$ along with its associated CWE type $T_i$, a Deep Learning (DL) model generates the corresponding repaired code $Y_i$.

\vspace{0.1cm} 
\noindent\textbf{Fusion-in-Decoder (FiD) Framework} is originally proposed for addressing challenges in the open-domain question-answering task in natural language processing (NLP)~\cite{FID}.
It is a sequence-to-sequence task, where the input is the question with numerous relevant passages and the output is the answer to the question. 
Before the introduction of FiD, researchers~\cite{lewis2020retrieval} concatenated the question and relevant passages to create the input for the encoder. 
However, the encoder had a limitation on the maximum input length, which led to the truncation of the input data (i.e., the concatenated questions and passages) and resulted in a substantial information loss.

The FiD framework adopts a divide-and-conquer approach to address such length limitation~\cite{FID}. 
Specifically, each question and passage is individually encoded using an encoder, producing contextual embeddings. These contextual embeddings are subsequently concatenated to form a composite representation of the questions and all passages, which is then fed into a decoder to generate the desired output. 
By incorporating the FiD framework, Transformer-based models surpass the prescribed length limitation and gain the ability to effectively comprehend and encode numerous passages.

%% file: Sections/3_approach.tex
\section{Approach}
\label{sec:aapporach}

\begin{figure*}[t] 
\centering 
\includegraphics[width=1\linewidth]{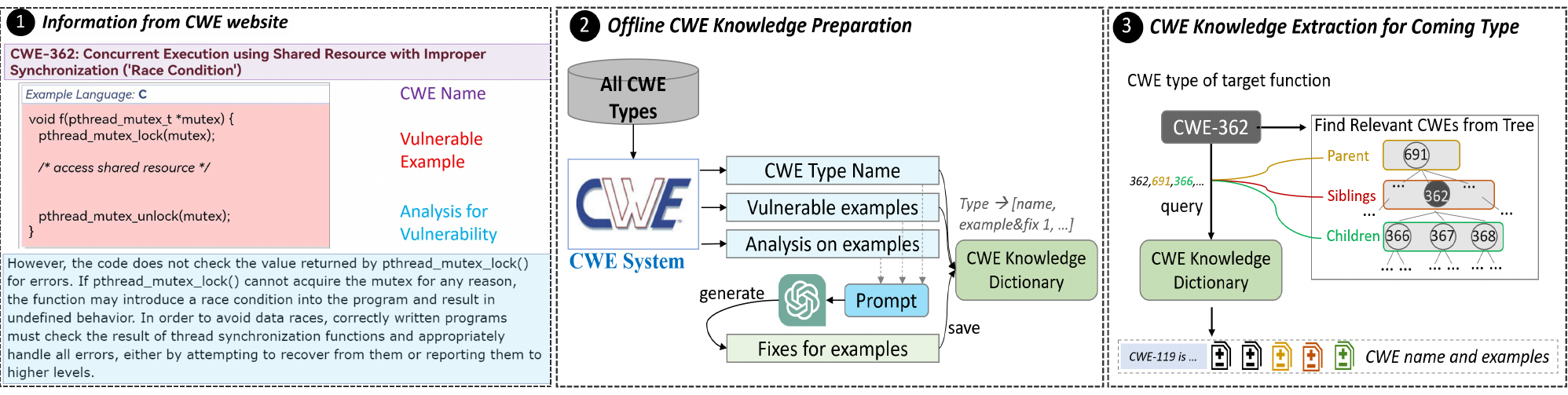} 
\vspace{-0.8cm}
\caption{Details of CWE knowledge extraction. \ding{182} the CWE name and one example from the CWE website; \ding{183} the process of generating fixes of typical vulnerable examples from CWE; \ding{183} the process to obtain vulnerable-fix code pairs given the target CWE.}
\vspace{-0.4cm}
\label{fig:cwe_extraction} 
\end{figure*}

The framework of VulMaster is presented in Figure~\ref{fig:framework}. VulMaster
takes a vulnerable function and its CWE type as input and generates the corresponding fix. 
VulMaster involves three main parts, where the first two parts prepare the input data for the VulMaster model, and the third part is responsible for encoding the inputs and generating the fixed code.

\vspace{0.04cm}
\noindent\textbf{Part 1: Vulnerable Code Preprocessing.} 
Given a vulnerable function $X_i$, this part preprocesses the function into a code token sequence and the AST node sequence by traversing the ASTs.

\vspace{0.04cm}
\noindent\textbf{Part 2: CWE knowledge Extraction.}
Given the CWE type $T_i$ of the vulnerable function $X_i$, this part outputs the name, vulnerable code examples listed on CWE web pages, and the corresponding fixes for vulnerable code examples from CWE web pages.

\vspace{0.04cm}
\noindent\textbf{Part 3: FiD and Relevance Prediction.}
This part first encodes all the input data (i.e., the code token sequences, AST node sequences, and CWE knowledge) into the contextual embeddings in a divide-and-conquer manner. 
Subsequently, the contextual embeddings are aggregated together.
Then, the aggregated embedding is fed into the decoder to generate the corresponding repair $Y_i$.

\vspace{0.04cm}
\noindent\textbf{Backbone model.}
VulMaster needs a pre-trained model as the backbone model and we employ CodeT5-base~\cite{CodeT5} for its great performance on sequence-to-sequence tasks~\cite{niu2023empirical}.

\vspace{-0.4cm}
\subsection{Vulnerable Code Preprocessing}
This part aims to perform preprocessing on the target vulnerable functions to extract code segments and AST node sequences.

\vspace{0.1cm}
\noindent\textbf{Tokenization.}
To correctly leverage the CodeT5 model, we utilize the CodeT5 tokenizer~\cite{codet5_huggingface} to tokenize each vulnerable function into a token sequence.

\vspace{0.1cm}
\noindent\textbf{Vulnerable function partitioning.}
VulMaster divides lengthy input code into multiple segments, ensuring that each segment adheres to the length limit (i.e., 512 tokens in the case of CodeT5). This helps VulMaster to further process and understand each of the code segments. Specifically, the process starts with tokenization, where the entire function is transformed into a sequence of tokens. VulMaster then proceeds to partition this token sequence into segments, ensuring that each segment contains no more than 512 tokens. If a function is shorter than 512 tokens, it constitutes a single segment.

\vspace{0.1cm}
\noindent\textbf{AST node sequences.}
Existing approaches~\cite{vrepair,vulrepair} treat the input code as unstructured natural language text, disregarding the rich structure inherent in source code that can provide valuable semantic information.
A notable structural representation of source code is the AST. 
Thus, we first adopt the {\tt tree-sitter} parser to transform the input vulnerable function into the AST representation.
However, pre-trained code models like CodeT5 are not designed to directly handle AST, as they are mainly optimized for sequential data rather than tree or graph structures.
To address this limitation, we adopt a depth-first traversal method to convert the AST into an AST node sequence while preserving the structural information~\cite{hu2018deep}.
To generate the AST node sequence, we start by adding the concatenation of the \textit{type} and \textit{value} of the root node to the empty list $A$. 
The \textit{value} refers to the actual token present in the source code, while the \textit{type} represents the AST node's type. 
Next, we traverse the sub-trees of the root node in a depth-first order, adding the concatenation of values and types of the root nodes of each sub-tree into the list $A$. This recursive process is repeated for each sub-tree until all nodes within the tree have been traversed.
Finally, the AST node sequence $A$ is divided into multiple segments, each adhering to the length limit of 512 tokens.

\vspace{-0.2cm}
\subsection{CWE knowledge Extraction}
This part involves gathering rich information from the CWE website for a CWE type of target vulnerable function.

\vspace{0.1cm}
\noindent\textbf{Offline CWE knowledge preparation.} 
To provide the CWE knowledge for an incoming vulnerable function instantly, we first build a CWE knowledge dictionary in an offline manner. The CWE knowledge dictionary is a dictionary where the key of each item is the CWE type and the corresponding value is a list of CWE knowledge, i.e., the CWE name, the vulnerable examples from CWE websites, and the fixes for those examples generated by ChatGPT.
Figure~\ref{fig:cwe_extraction} (\ding{182}) showcases the information available on the website of CWE-362, including the CWE name, the vulnerable examples from CWE websites, and the analysis of the examples written by experts. 
Note that the CWE website usually only provides vulnerable examples without fixed examples. 
Although vulnerable code examples exist, they do not possess the essential knowledge on ``how to fix vulnerabilities''.
The analyses of the vulnerable examples contain the ``how to fix'' knowledge, but they are in natural language texts rather than source code.
Therefore, we seek suitable fixes for the vulnerable code examples to complement the ``how to fix'' knowledge. To obtain the fixed code examples, we adopt the recent advancements in generative AI models, i.e., ChatGPT, which is proficient in generating code given detailed guidance~\cite{chatgpt,wang2023codet5+}.

Figure~\ref{fig:cwe_extraction} (\ding{183}) presents the overall process of building this offline CWE knowledge dictionary for VulMaster.
Firstly, we begin by gathering all the CWE types that appeared in the experiment dataset~\cite{vulrepair}. 
Secondly, for each CWE type, we access its corresponding website to extract valuable information, i.e., the CWE name, typical vulnerable examples, and expert analysis of these examples. 
Thirdly, for each vulnerable code example and its accompanying analysis, we leverage ChatGPT to generate potential fixes. 
Specifically, we crafted the following prompt to generate the fixes:
\vspace{-0.1cm}
\begin{center}
\begin{tcolorbox}[colback={blue!5!white},
                  colframe=black,
                  width=8.5cm,
                  arc=1mm, auto outer arc,
                  boxrule=0.2pt,
                  top=0pt,
                  bottom=0pt
                 ]

``The code $\{code\}$ contains a vulnerability of type $\{name\}$. The analysis of this vulnerable code is $\{analysis\}$. Please generate the repaired code to address the vulnerability:''
\end{tcolorbox}
\end{center}
\vspace{-0.1cm}
The $\{code\}$, $\{name\}$, and $\{analysis\}$ are respectively filled in the vulnerable code example, CWE name, and expert analyses on this vulnerable example listed on the CWE web page. 
We utilize the prompt for each vulnerable example with its name and expert analysis and query ChatGPT (i.e., the {\tt GPT-3.5-turbo} model with its default setting~\cite{chatgpt}) to obtain the fixed code. 
Given the ample and accurate expert analysis as guidance and the simplicity of those vulnerable examples from CWE, ChatGPT could generate a large number of correct fixes for those vulnerable examples from CWE web pages. 
We discuss such performance of ChatGPT for vulnerable code examples from CWE in Section~6.2.

However, different from the vulnerable code examples in CWE web pages, vulnerabilities in real-world software~\cite{vulrepair} often lack expert analyses, and the complexity of vulnerabilities in real-world software is significantly increased. 
Such factors make ChatGPT struggle to generate accurate fixes for real-world vulnerabilities, which is also discussed in Section~5.1.

\vspace{0.1cm}
\noindent\textbf{CWE knowledge extraction for coming data.}
Given the CWE type of one target vulnerable function, VulMaster can easily access the CWE knowledge, including the CWE name, vulnerable code examples, and the generated fixes, by querying the CWE knowledge dictionary.
In addition to extracting vulnerable examples and fixes for the target CWE type, VulMaster goes beyond this and includes examples and fixes from other closely related CWE types in the input. This process is represented as \ding{184} in Figure~\ref{fig:cwe_extraction}.
The decision to include examples and fixes from closely related CWE types stems from one observation: CWE types are organized hierarchically, with parent, child, and sibling types sharing many similarities with a specific CWE type. 
For instance, CWE-125 represents ``Out-of-bounds Read'', while its children, such as ``CWE-126: Buffer Over-read'' and ``CWE-127: Buffer Under-read'', are more specific cases of CWE-125.
Thus, VulMaster can consider a broader range of relevant code examples and fixes by incorporating examples and fixes from parent, child, and sibling CWE types (i.e., the wide range) in the input, rather than solely relying on those from the input CWE type (i.e., the narrow range).

\vspace{-0.2cm}
\subsection{Fusion-in-Decoder and Relevance Prediction}

\noindent\textbf{Backbone model adaption for AST.}
Although CodeT5 is pre-trained on a large number of source code snippets, it has not seen AST node sequences during its pre-training, resulting in a lack of familiarity with such structural representation.
Prior studies~\cite{zhou2021assessing,zhou2023ccbert,he2023representation} have suggested that the lack of pertinent software artifacts during pre-training may lead to suboptimal pre-trained models, as exemplified in pre-trained models for code changes~\cite{zhou2023ccbert}, Android bytecode~\cite{sun2023dexbert}, and Stack Overflow posts~\cite{he2023representation}. A straightforward improvement involves incorporating the corresponding data during (additional) pre-training.
Following this idea, we introduce an additional pre-training step to enhance the backbone model's understanding of AST structures, which requires a substantial and relevant code corpus. 
We choose to use the bug-fixing corpus provided by Chen et al.~\cite{vrepair}, consisting of over 500,000 pairs of buggy and fixed functions. 
The bug-fixing task shares similarities with vulnerability repair~\cite{vrepair}, making this corpus a valuable resource for us. 
Specifically, we further pre-train the CodeT5 model to generate the fixed version of buggy code. 
For half of the training samples, the inputs are AST node sequences of buggy code, while for the other half are the buggy code itself. 
Both the lengthy AST node sequences and source code are truncated to the first 512 tokens for simplicity.
We also include the source code as a part of the inputs to ensure that the trained CodeT5 retains its ability to understand both the source code and AST semantics.
In this model adaption phase, we employ the Adam optimizer~\cite{adam} with a learning rate of 2e--5 and train CodeT5 for 5 epochs on the bug-fixing corpus.
We denote the model after this step as \textit{adapted-CodeT5}.

\vspace{0.1cm}
\noindent\textbf{Contextual Embedding.}
As shown in Figure~\ref{fig:framework}, VulMaster uses the encoder model to generate contextual embedding for the following input sources: 1) the complete vulnerable function $X_i$, 2) AST node sequences, 3) CWE type name, 4) examples and corresponding fixes of the CWE type $T_i$ and related CWE types.
Specifically, for one pair of vulnerable code example and its fix, we concatenate them into a single sequence and feed it to the encoder to get the contextual embedding.
Notably, the encoder utilized is the encoder model of the \textit{adapted-CodeT5}.

\vspace{0.1cm}
\noindent\textbf{Relevance prediction.}
VulMaster is equipped with vulnerable-fixed code pairs gathered from web pages related to the input CWE type  $T_i$ (i.e., the CWE type of the input vulnerable function in the evaluation dataset) and its associated parent/child/sibling CWE types. However, not all vulnerable-fixed code pairs from CWE web pages hold the same level of relevance when it comes to repairing the input vulnerable function in the evaluation dataset~\cite{cvefixes,fan2020ac}. Pairs associated with the input CWE type are deemed the most relevant, while those linked to the input CWE type's parent/child/sibling CWE types are considered less relevant. The objective of the relevance prediction module is to assist VulMaster in identifying and highlighting the most relevant vulnerable-fixed code pairs from CWE web pages, while also considering less relevant pairs as contextual information. To achieve this, we introduce explicit supervision by performing a binary classification task~\cite{wang2023rfid}.

Then we explain the preparation of the input for the relevance prediction module. The input is constructed by the following steps: Firstly, we collect vulnerable code snippets from the CWE web pages of the input CWE type $T_i$ and its parent/child/sibling CWE types. Secondly, we collect the respective fixes (generated by ChatGPT) for the vulnerable code snippets from the first step. Thirdly, we concatenate each vulnerable code snippet from the first step with its corresponding fix, creating a single sequence. The resulting concatenated sequences (representing vulnerable-fixed code pairs from CWE web pages) are the input to the relevance prediction module.
For the outputs (labels),  we label each vulnerable-fixed code pair from CWE web pages as ``most related'' (class 1) if it belongs to the input CWE type $T_i$.  Otherwise, it is labeled as ``less related'' (class 0).

The k-th vulnerable-fixed code pair from CWE webpages is transformed into its embedding $E_k$ by the encoder model of VulMaster. Then the contextual embedding $E_k$ is fed to a binary classifier (i.e., MLP~\cite{hastie2009elements}) to predict whether this k-th pair is ``most related'' or ``less related''. The output of the binary classifier is denoted as $p_k = \text{Classifier}(E_k)$. To update the model, we minimize the Cross-Entropy loss:  
$L_{relevance} = \sum_{k=1}^{K} -{(g_k\log(p_k) + (1 - g_k)\log(1 - p_k))}$.
Here, $g_k$ represents the ground truth relevance label of the $k$-th vulnerable-fixed code pair from CWE webpages and $p_k$ is the prediction score of the pair.
By optimizing this loss, a model can learn to effectively distinguish between ``most related'' and ``less related'' vulnerable-fixed code pairs from CWE webpages.

\vspace{0.1cm}
\noindent\textbf{Fusion-in-Decoder.}
VulMaster combines all contextual embeddings of the input components into a single concatenated embedding, denoted as $C_{encoder}$. 
The concatenation is performed as follows: 
$C_{encoder} = [I_1, ... , I_n; A_1, ... A_m; D; E_1,...,E_k]$
where ``$;$'' indicates the concatenation operation. 
$I_j$ represents the $j$-th segment of the input vulnerable function, $A_j$ represents the $j$-th segment of the AST node sequence of the input vulnerable function, $D$ represents the CWE name and $E_j$ represents the $j$-th example and its fix from CWE.
The concatenated contextual embedding $C_{encoder}$ is then passed into the decoder model (i.e.,  the decoder of the \textit{adapted-CodeT5}) to generate the fixed code. 
In general, the model is updated to minimize the loss: 
$L_{repair} = -log  \; p (Y_{i} | X_{i}, AST_{i}, Name_{i}, \\ \; \; \; Example\text{-}Fix\text{-}Pairs_{i}  )$
This minimization aims to increase the probability of the model generating the correct fixed function ($Y_{i}$) using the provided diverse input components.

\vspace{0.1cm}
\noindent\textbf{Multi-task Learning.}
Previous research has demonstrated that multi-task learning is beneficial in enhancing the performance of learning-based models in tasks such as code completion~\cite{Liu_Li_Wei_Xia_Fu_Jin_2022}, code generation ~\cite{Wang_Liu_Zhou_Liu_Liu_Wu_Cui_2022}, and code understanding~\cite{Wang_Yu_Li_Dong_Wang_Qing_2021}. In the case of VulMaster, the adoption of the multi-task learning framework also aims to improve its effectiveness. Specifically, VulMaster is simultaneously trained on two tasks: 1) the relevance prediction task which identifies the most relevant vulnerable-fixed code pairs from CWE web pages from a pool of relevant pairs (i.e., $L_{relevance}$); 2) the vulnerability repair task which generates the fix for the input vulnerable function from the evaluation datasets~\cite{cvefixes,fan2020ac} (i.e., $L_{repair}$). The total loss function is the sum of the losses from each task: $L = L_{relevance} + L_{repair}$.

%% file: Sections/4_setup.tex
\section{Experimental Design}
\label{sec:setup_and_results}

\subsection{Dataset for Evaluation}
We utilize the same real-world C/C++ vulnerability dataset that is used to evaluate the existing vulnerability repair approaches (i.e., VulRepair~\cite{vulrepair} and VRepair~\cite{vrepair}).
It consists of 8,482 pairs of vulnerable C/C++ functions and their corresponding fixes by merging two existing datasets: CVEFixes~\cite{cvefixes} and Big-Vul~\cite{fan2020ac}. 
Please note that VulMaster is evaluated using the identical dataset combination of CVEFixes and Big-Vul, which was previously employed by the learning-based state-of-the-art VulRepair~\cite{vulrepair}.
These pairs are collected from 1,754 open-source software projects spanning the period from 1999 to 2021. 
This dataset is divided into training (70\%), testing (20\%), and validation (10\%) subsets by previous studies~\cite{vulrepair}.
The data statistics are shown in Table~\ref{tab:statistics}. 
Notably, we checked and confirmed that the vulnerable-fixed code pairs sourced from the CWE website (Section 3.2) have no duplicates with the vulnerability repair evaluation dataset~\cite{vulrepair} mentioned above.

\vspace{0.08cm}
\noindent\textbf{Processing.}
To process the dataset, VulRepair~\cite{vulrepair} and VRepair~\cite{vrepair} adopt the same steps, which involve the addition of special tokens.
Particularly, each vulnerable function in the dataset is marked using the special tokens \emph{<StartLoc>} and \emph{<EndLoc>}. 
The \emph{<StartLoc>} token indicates the beginning of the vulnerable code lines, while the \emph{<EndLoc>} token indicates the end.
For the ground truth, the special tokens \emph{<ModStart>} and \emph{<ModEnd>} are inserted into each repaired function to signify the start and the end of the repaired code lines, respectively. 
These special tokens serve the purpose of guiding the model's attention toward the vulnerable code lines and the corresponding fixed code lines. 
In other words, \emph{<StartLoc>} and \emph{<EndLoc>} provide the \textit{perfect vulnerability localization} to vulnerability repair approaches.

\begin{table}[t]
\centering
\caption{Statistics of the studied dataset}
\label{tab:statistics}
\vspace{-0.3cm}
\resizebox{0.47\textwidth}{!}{%
\begin{tabular}{l|c|c|c|c|c|c}
\hline
\textbf{Dataset} & \textbf{Train} & \textbf{Valid} & \textbf{Test} & \textbf{All} & \multicolumn{1}{c|}{\textbf{\begin{tabular}[c]{@{}c@{}}\%samples\\ \textgreater 512 tokens\end{tabular}}} & \multicolumn{1}{c}{\textbf{\begin{tabular}[c]{@{}c@{}}\%samples\\ \textgreater{}1000 tokens\end{tabular}}} \\ \hline
Original         & 5,937          & 839            & 1,706         & 8,482        &   45.6\%                                                                                                       &  25.0\%                                                                                                          \\ \hline
Deduplication    & 3,872               &  316              & 1,612         & 5,800        &   44.9\%                                                                                                        &   25.1\%                                                                                                         \\ \hline
\end{tabular}
}
\vspace{-0.4cm}
\end{table}

\vspace{0.08cm}
\noindent\textbf{Label leakage issue.}
In this paper, the reported performance of the learning-based SOTA VulRepir is different from its original paper~\cite{vulrepair}. This is caused by our leakage correction.
The dataset used in VulMaster is merged from two existing datasets CVEFixes~\cite{cvefixes} and Big-Vul~\cite{fan2020ac}. 
The authors of VulRepair seem to assume that data samples from two datasets are different from each other. 
By checking the vulnerable and fixed code pairs from each dataset, we found that about 60\% of the pairs are duplicates. 
As the authors of VulRepair~\cite{vulrepair} further randomly split the merged dataset into training/validation/test sets, it leads to the presence of duplicated samples across the training, validation, and test sets.
The label leakage issue might lead to an overestimation in performance measurement and artificially inflated scores~\cite{allamanis2019adverse}.
To tackle this problem, we conducted a deduplication operation in three steps. Firstly, we removed any samples from the training set that were identical to any samples in the validation or test sets. 
Next, we eliminated any samples from the validation set that were identical to any samples in the test set. 
Finally, we eliminated 94 duplicate samples within the test set.
The deduplication statistics are presented in Table~\ref{tab:statistics}.

To assess the impact of the label leakage issue, we first use the replication package~\cite{vulrepair_rep} released by the authors of VulRepair to train VulRepair from scratch by using the original fine-tuning dataset.
Experimental results indicate just a 0.2\% difference between our replication and the reported performance in the VulRepair paper, validating the correct usage of their replication package. Next, we trained another VulRepair from scratch on the deduplicated version of the dataset, resulting in a decrease in VulRepair's performance from 44.2\% to 10.2\% in terms of Exact Match accuracy. 
This significant drop is expected because of the label leakage issue.

\vspace{-0.2cm}
\subsection{Baselines}
We evaluate VulMaster against three groups of baseline models. The first group consists of learning-based automatic vulnerability repair approaches, namely VulRepair~\cite{vulrepair} and VRepair~\cite{vrepair}.
Both of these approaches take a vulnerable function concatenated with a CWE ID (e.g., CWE-119) as input and generate the fixed code as output. 
The second group comprises general-purpose pre-trained code models that are widely used in various code-related downstream tasks. 
Specifically, we include two encoder-based models (i.e., CodeBERT~\cite{CodeBERT} and GraphCodeBERT~\cite{GraphCodeBERT}), two decoder-based models (i.e., PolyCoder-160M~\cite{polycoder} and CodeGen-350M~\cite{codegen}), and two encode-decoder based models (i.e., CodeReviewer~\cite{codereviewr} and CodeT5-base~\cite{CodeT5}).
Those models take the same input and output as VulRepair and VRepair.
The third group consists of large language models (LLMs) for code, which have gained significant attention recently~\cite{xia2023automated,kang2023large,ahmed2023recommending}.
For LLMs, we include the well-known ChatGPT model (i.e., gpt-3.5-turbo)~\cite{chatgpt} and the improved version of ChatGPT model (i.e., gpt-4)~\cite{gpt4} as baselines. 
We use a prompt similar to the prompt described in Section 3.2. We remove the sentence about expert analysis because of the lack of expert analyses in real-world vulnerabilities. 
Additionally, one sentence is added to explicitly tell LLMs about the meaning of special tokens (i.e.,  \emph{<StartLoc>} and \emph{<EndLoc>}). 
We also provide the CWE description and one vulnerable example from the CWE website in the prompt. The prompt used is:
\textit{``A vulnerability of type $\{cwe\_type\}$ refers to $\{cwe\_description\}$. One vulnerable example of this type is $\{cwe\_example\}$. The code $\{code\}$ contains a vulnerability of type $\{cwe\_type\}$. Note that <StartLoc> and <EndLoc> indicate the start and the end of vulnerable code lines. Please generate the repaired code to address the vulnerability:''}.

\vspace{-0.2cm}
\subsection{Experimental Setting}

\noindent\textbf{Implementation details.}
Following prior work~\cite{vulrepair,vrepair}, we also focus on fixing C/C++ vulnerabilities. Thus, we mainly collected C/C++ vulnerable examples from the CWE websites. But if there is no C/C++ vulnerable example for a CWE type, we will also collect one example of other languages like C\# and Java. 
Please note that our approach is generic and language-agnostic and can be applied to any programming language.
During training, we employ the Adam optimizer with a learning rate of $1e-4$. The weight decay rate is set to 0.01. The batch size is set to 64, and the training process consists of 20 epochs. After each epoch, we evaluate the model's performance on the validation set. The best checkpoint on the validation set is selected for the final testing.
For the hyper-parameters of the FiD framework, the maximum length of an individual segment (e.g., a segment of a lengthy vulnerable function) is set to 512, aligning with the maximum length of CodeT5~\cite{CodeT5}. 
Additionally, we observe that about 90\% of the samples in our dataset comprise fewer than 10 segments, each containing up to 512 tokens. These segments encompass various input components, including the vulnerable function, AST node sequences, CWE descriptions, and examples from relevant CWE types. Therefore, we set the maximum number of segments of FiD to 10. 
We will discuss the impact of the choice of the maximum number of segments in Section 5.3 (RQ3).

\vspace{0.08cm}
\noindent\textbf{Evaluation metric.}
In accordance with previous studies~\cite{vulrepair}, we employ the Exact Match (EM) metric as our evaluation measure. EM refers to the percentage of the generated code that has the same token sequence as the ground truth.
We also utilize another widely used metric, i.e. BLEU-4~\cite{bleu} score, which evaluates the token-level similarity between the generated code and the ground truth.
In addition, we utilize the CodeBLEU~\cite{codebleu} score, which is a specialized variant of the BLEU score tailored for source code by additionally taking code structure into account.
We consider the \textit{Top 1 prediction} when calculating metrics.

%% file: Sections/5_result.tex
\vspace{-0.2cm}
\section{Experimental Results}
\label{section:results}

Our work aims to answer three research questions (RQ).  

\begin{itemize}[leftmargin=*]
    \item \textbf{RQ1: How effective is VulMaster compared to baselines?} In RQ1, we compare our VulMaster to learning-based SOTA vulnerability repair approaches, widely used pre-trained code models, and the latest LLMs.
    \item \textbf{RQ2: How do the key designs of VulMaster influence the model performance?} In RQ2, we conduct an ablation study to confirm the contributions of different modules.
    \item \textbf{RQ3: What are the influences of different design choices?} In RQ3, we investigate how different design choices impact the effectiveness of our VulMaster.
\end{itemize}

\input{Sections/RQ1}

\input{Sections/RQ2}

\input{Sections/RQ3}

%% file: Sections/RQ1.tex
\vspace{-0.2cm}
\subsection{RQ1. The Effectiveness of VulMaster}

\begin{table}[t]
\centering
\caption{Model performance of baselines and VulMaster} \label{tab:RQ1}
\vspace{-0.3cm}
\resizebox{1\columnwidth}{!}{%
\begin{tabular}{@{}llrrr@{}}
\toprule
\textbf{Type} & \multicolumn{1}{c}{\textbf{Approach}} & \textbf{EM} & \textbf{BLEU} & \multicolumn{1}{r}{\textbf{CodeBLEU}} \\ \midrule
\multirow{1}{*}{\textbf{pre-trained}} & CodeBERT~\cite{CodeBERT}                 &3.9   & 3.9 &12.2 \\
                                &\textit{+ bug-fixing and CWE data}   &7.3   &6.3  &22.0 \\ \cdashline{2-5}
                                      & GraphCodeBERT~\cite{GraphCodeBERT}       &3.6  &2.0   &9.9  \\
                                &\textit{+ bug-fixing and CWE data}  &8.1  &5.2   &16.7 \\ \cdashline{2-5}
                                      & PolyCoder~\cite{polycoder}                 &3.5  &4.3  & 9.9 \\
                                 &\textit{+ bug-fixing and CWE data}   &9.9  &14.9  &30.4 \\ \cdashline{2-5}
                                      & CodeGen~\cite{codegen}             &7.0  & 4.7 & 12.1 \\ 
                                 &\textit{+ bug-fixing and CWE data}   &12.2  &17.4 &30.3  \\ \cdashline{2-5}
                                      & Codereviewer~\cite{codereviewr}        &7.3  &12.0 &33.5 \\
                                  &\textit{+ bug-fixing and CWE data}    &10.2  &13.0 &38.0 \\ \cdashline{2-5}
                                      & CodeT5~\cite{CodeT5}          &10.2   &21.3  &32.5  \\ 
                                  &\textit{+ bug-fixing and CWE data}   &16.8   &24.2  &35.3\\ \midrule
\multirow{1}{*}{\textbf{LLM}}         & GPT-3.5~\cite{chatgpt}             &3.6  & 8.8 & 17.6\\
                                      & GPT-4~\cite{gpt4}             &5.3  & 9.7 &16.6  \\
                                      \midrule
\multirow{1}{*}{\textbf{task-specific}}        
                                 & VRepair~\cite{vrepair}             &8.9   &11.3  &31.8 \\
                                 &\textit{+ bug-fixing and CWE data}   &8.9   &11.4  &31.7 \\ \cdashline{2-5}
                                 & VulRepair~\cite{vulrepair}  (SOTA)         &10.2   &21.3  &32.5  \\
                                 & \textit{+ bug-fixing and CWE data}   &16.8   &24.2  &35.3  \\  \midrule
\textbf{Ours}                    & VulMaster           & \textbf{20.0}  &\textbf{29.3} &\textbf{40.9} \\ \bottomrule
\end{tabular}
}
\vspace{-0.4cm}
\end{table}

\noindent
\textbf{Setup.} We evaluate both the baselines (described in Section 4.2) and our VulMaster model using the dataset under investigation (discussed in Section 4.1). The evaluation metrics, namely EM, BLEU, and CodeBLEU, are detailed in Section 4.3. Notably, higher scores for all these metrics indicate better performance.

Moreover, in the development of VulMaster, we employed a bug-fixing dataset from~\cite{vrepair} and vulnerable-fixed code pairs obtained from the CWE systems. However, the baseline models had not previously been fine-tuned on the same bug-fixing corpus or CWE data, potentially resulting in an unfair comparison. To mitigate this potential bias, we conducted additional experiments with the baselines, aligning them with our approach. This involved a four-step process: 
Firstly, we merged the bug-fixing dataset~\cite{vrepair} with the vulnerable-fixed code pairs from the CWE system, creating a merged single dataset. Secondly, we conducted fine-tuning on this merged dataset for the baselines (with the exception of close-sourced GPT-3.5 and GPT-4). The objective of this fine-tuning process was to generate clean code snippets given the presence of buggy or vulnerable code. Thirdly, the baselines were further fine-tuned using the vulnerability repair dataset~\cite{cvefixes, fan2020ac} to prepare them for the vulnerability repair task. Finally, the baselines generated predictions for the test set of the vulnerability repair dataset. The results achieved after enhancing the bug-fixing and CWE data in this manner are denoted by the label \textit{``+bug-fixing and CWE data''}.

\vspace{0.1cm}
\noindent
\textbf{Results.}
Table~\ref{tab:RQ1} presents the performance comparisons between VulMaster and the baseline models, with the best results highlighted in bold. \textit{\textbf{Our VulMaster achieves the best results among all baselines and outperforms the learning-based SOTA by a large margin.}}
The experimental results illustrate that VulMaster enhances the EM, BLEU, and CodeBLEU scores from 10.2\% to 20.0\%, 21.3\% to 29.3\%, and 32.5\% to 40.9\%, respectively, compared to the state-of-the-art learning-based vulnerability repair approach VulRepair.
We conducted the Wilcoxon signed-rank test~\cite{wilcoxon1992individual} on the paired data between VulMaster and each of the baseline models. 
The resulting p-values for all the comparisons were found to be less than 0.001, indicating that the performance differences between VulMaster and the baseline models are statistically significant.

Furthermore, LLMs such as GPT-3.5 and GPT-4 struggle to effectively repair real-world software vulnerabilities possibly because they cannot update their parameters to fully utilize the training data. Sun et al.~\cite{sun2023automatic} also reported a similar phenomenon in code summarization, where GPT-3.5 performs notably worse than fine-tuned pre-trained CodeT5 in terms of BLEU scores.

Table~\ref{tab:RQ1} also highlights the improvements achieved by incorporating the bug-fixing corpus and CWE data into all baselines. Despite these improvements, VulMaster maintains a substantial lead over all baselines. Notably, VulMaster outperforms the best-performing baseline (the enhanced VulRepair) by a significant margin, with improvements of 19.0\%, 21.1\%, and 15.9\% in terms of EM, BLEU, and CodeBLEU, respectively. One of the novel aspects of VulMaster is the usage of CWE data that was not leveraged in prior works. These experiment results demonstrate that the reason behind VulMaster’s superior performance over the baselines is not the inclusion of the CWE data alone. Rather, the other unique designs of VulMaster, such as 1) the Fusion-in-Decoder module that handles lengthy vulnerable functions, and 2) the AST node sequence that provides vulnerable code structures to the model, are also beneficial.

\begin{figure}[t] 
\centering 
\includegraphics[width=0.6\linewidth]{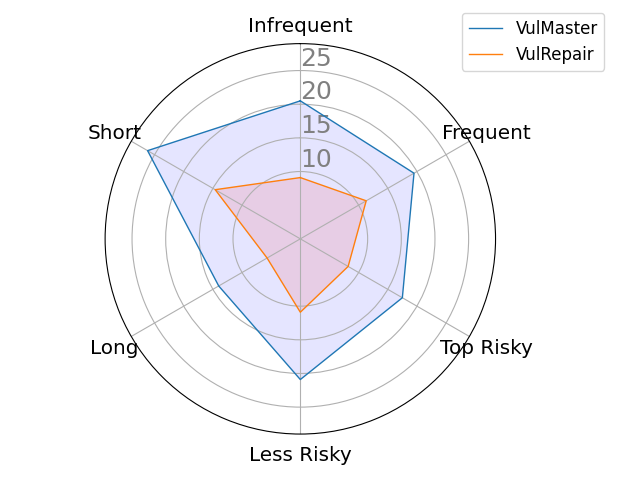} 
\vspace{-0.4cm}
\caption{Model performance in EM on different groups.}
\vspace{-0.4cm}
\label{fig:radar} 
\end{figure}

\vspace{0.1cm}
\noindent
\textbf{Analyses.}
We further analyze the learning-based SOTA VulRepair (also the best-performing baseline) and our VulMaster. Specifically, we divided the whole test set into subgroups and observed how VulRepair and VulMaster perform in different subgroups. Those subgroups are split based on three different criteria: the lengths of the input functions, the frequencies of vulnerabilities, and the severity levels of vulnerabilities. In terms of the lengths of the input, two subgroups were created: \textit{1) the long function group} consisted of half of the test samples whose vulnerable function lengths are longer, and \textit{2) the short function group}, consisted of the remaining test samples. The threshold between long/short groups is 449 tokens.
In terms of the frequencies of vulnerabilities, we have: \textit{3) the frequent group}, consisted of 50\% of the test samples whose CWE types are more frequent, and \textit{4) the infrequent group}, consisted of the remaining half of test samples. 
For the severity levels of vulnerabilities, we have: \textit{5) the top risky group} consisted of test samples whose CWE types are one of the top 10 most dangerous CWE types, and \textit{6) the less risky group} consisted of the remaining test samples.

Figure~\ref{fig:radar} shows the EM performance for each subgroup. \textit{\textbf{VulMaster exhibits significant superiority over the learning-based SOTA in all investigated dimensions.}} For example, VulMaster could improve the EM scores from 8.2\% to 17.5\% for the top risk group. Table~\ref{tab:dangerous} presents the number of perfect predictions made by VulMaster for the top risk group. 
Although the correct fix ratio (17.5\%) of VulMaster is not high enough, this improvement over the SOTA is significant and suggests that our method is a substantial step towards more practical automatic vulnerability approaches in the future.
Additionally, we observe that VulRepair exhibits lower performance with infrequent vulnerabilities compared to frequent ones, consistent with findings from a recent study~\cite{zhout2023devil} that suggested learning-based approaches may struggle with infrequent data. In contrast, VulMaster achieves comparable performance across both frequent and infrequent vulnerabilities, demonstrating its great performance with less common data.

\begin{table}[t]
\centering
\caption{Numbers of perfect predictions for the top-10 most dangerous CWEs} \label{tab:dangerous}
\vspace{-0.3cm}
\resizebox{1\columnwidth}{!}{%
\begin{tabular}{@{}cllrrr@{}}
\toprule
\textbf{Rank} &
  \multicolumn{1}{c}{\textbf{CWE-Type}} &
  \multicolumn{1}{c}{\textbf{Name}} &
  \multicolumn{1}{c}{\textbf{VulRepair}} &
  \multicolumn{1}{c}{\textbf{VulMaster}} &
  \multicolumn{1}{c}{\textbf{\#sample}} \\ \midrule
1                    & CWE-787 & Out-of-bounds Write        &3   &12   & 58  \\
2                    & CWE-79  & Cross-site Scripting       &0   &1   & 1   \\
3                    & CWE-89  & SQL Injection              &0   &1   & 4   \\
4                    & CWE-416 & Use After Free             &1   &6   & 60  \\
5                    & CWE-78  & OS Command Injection       &0   &0   & 4   \\
6                    & CWE-20  & Improper Input Validation  &18   &33   & 128 \\
7                    & CWE-125 & Out-of-bounds Read         &12   &20   & 156 \\
8                    & CWE-22  & Path Traversal             &0   &0   & 5   \\
9                    & CWE-352 & Cross-Site Request Forgery &0   &0   & 1   \\
10                   & CWE-434 & Dangerous File Type        & - & - & 0   \\ \midrule
\multicolumn{1}{l}{} &         & \multicolumn{1}{r}{TOTAL}  &34 (8.2\%)   &73 (17.5\%)   & 417 \\ \bottomrule
\end{tabular}
}
\vspace{-0.4cm}
\end{table}

\vspace{0.2cm}
\noindent
\begin{tcolorbox} [boxrule=0.8pt,
                top=0.2pt,
                  bottom=0.2pt]
    \textbf{Answer to RQ1}: 
    VulMaster achieves the best performance (20.0\% in EM) among all baselines. 
    Specifically, VulMaster enhances the EM, BLEU, and CodeBLEU scores from 10.2\% to 20.0\%, 21.3\% to 29.3\%, and 32.5\% to 40.9\%, compared to the state-of-the-art learning-based approach.
\end{tcolorbox}

%% file: Sections/RQ2.tex
\vspace{-0.2cm}
\subsection{RQ2. The Impact of the Core of VulMaster}

\noindent\textbf{Setup.}
This RQ aims to investigate the contributions of the key designs of our approach.
Specifically, we conduct two sets of ablation studies using different backbone models: the input component designs and the DL model designs. 
In each ablation study, we remove one component at a time to examine the individual contributions of key designs.
\textit{(1) VulMaster w/o entire function}: removing the extended part of lengthy functions and only including the first 512 tokens, similar to the prior approaches like VulRepair,
\textit{(2) VulMaster w/o AST}: excluding the AST traversals, and
\textit{(3) VulMaster w/o CWE knowledge}: eliminating the CWE name, vulnerable examples from CWE, and generated fixes for CWE vulnerable examples by ChatGPT. 
For DL model designs, we experiment with three variants: 
\textit{(4) VulMaster w/o relevance-classifier}: disregarding the binary classification task used to identify relevant vulnerable examples,
\textit{(5) VulMaster w/o FiD}: removing the FiD, resulting in the simple utilization of the backbone model (truncating inputs at the 512 tokens).
\textit{(6) VulMaster w/o model adaptation}: leveraging the original CodeT5-base rather than the \textit{adapted-CodeT5} in Section 3.3.

\begin{table}[t]
\centering
\caption{Ablation study on VulMaster} \label{tab:RQ2}
\vspace{-0.3cm}
\resizebox{0.75\columnwidth}{!}{%
\begin{tabular}{@{}llr@{}}
\toprule
\textbf{Type}                     & \textbf{Variants}             & \textbf{EM}    \\ \midrule
     Full Model                   & VulMaster                     &  \textbf{20.0}         \\ \midrule
\multirow{1}{*}{Input Components} & \hspace{0.2em} -w/o entire function  &      18.8        \\
                                  & \hspace{0.2em} -w/o AST     &      19.0           \\
                                  & \hspace{0.2em} -w/o CWE knowledge     &        18.9     \\
                                  \midrule
\multirow{1}{*}{Model Designs}    & \hspace{0.2em} -w/o relevance prediction     & 19.6           \\
                                  & \hspace{0.2em} -w/o FiD                       &16.7   \\ 
                                  & \hspace{0.2em} -w/o model adaptation   & \underline{13.6}                     \\
\bottomrule
\end{tabular}
}
\vspace{-0.4cm}
\end{table}

\vspace{0.1cm}
\noindent\textbf{Results and Analyses.}
The experimental results are presented in Table~\ref{tab:RQ2}, with the best results highlighted in bold and the largest performance drop in the underline.
Our main findings are:
\textit{\textbf{1) All key designs are essential to achieve the best performance.}}
Based on the results shown in Table~\ref{tab:RQ2}, we observe that the absence of each key design in VulMaster leads to a reduction in the Exact Match score. It demonstrates that each component plays an important role in VulMaster. Specifically, the complete function, structural information, and CWE knowledge bring 6\%, 5\%, and 5.5\% improvements, respectively. For DL model designs, the auxiliary relevance prediction task, the FiD, and the model adaptation bring 2\%, 16.5\%, and 32\% improvements, respectively.
\textit{\textbf{2) The model adaptation is the most effective module.}} We find that incorporating model adaptation with the bug-fixing corpus contributes to a significant improvement in VulMaster. While Chen et al.~\cite{vrepair} demonstrated the effectiveness of bug-fixing corpus in the vanilla Transformer model, we are the first to show its high efficacy in adapting pre-trained models like CodeT5.

\vspace{0.2cm}
\noindent
\begin{tcolorbox} [boxrule=0.8pt,
                top=0.2pt,
                  bottom=0.2pt]
    \textbf{Answer to RQ2}: 
    All designs are essential for the performance of VulMaster. Besides, our input and model designs are effective and improve the performance by at most 32\% in EM. 
\end{tcolorbox}

%% file: Sections/RQ3.tex
\vspace{-0.2cm}
\subsection{RQ3. Influences of Design Choices}

\noindent\textbf{Setup.}
In this RQ, we experiment VulMaster with three main design choices: 1) the maximum number of segments, 2) the prompt used to generate fixes for vulnerable examples from the CWE system, and 3) the ChatGPT models used to generate the fixes for vulnerable examples.
For the maximum number of segments (denoted as $K$), it is a key hyperparameter in FiD~\cite{FID}, which controls the improved range of inputs. For instance, if $K=10$, then the maximum input range of VulMaster is broadened from 512 to 5,120 tokens. 
In this RQ, we conduct experiments by varying $K$ from 1 to 20, with an increment of 5. This aims to validate our choice of $K=10$.
For prompts, they can influence the performance of LLMs, like ChatGPT, to some extent~\cite{gpt3,t0}. While the prompt used in Section 3.2 may not be the optimal choice, it is impractical to explore all potential prompts~\cite{t0}. To examine VulMaster's sensitivity to different prompts, we evaluate three distinct prompts, as shown in Table~\ref{tab:prompts}. P0 is the prompt we used in Section 3.2, P1 is the rephrased version of P0, and P2 is a simplified version of P0. This experiment aims to demonstrate that VulMaster can deliver satisfactory performance even with varied prompts.
For ChatGPT models, OpenAI continuously and episodically updates its ChatGPT models~\cite{chatgpt}. 
To assess the performance of VulMaster to different versions of ChatGPT models, we experiment with three ChatGPT model versions: GPT-3.5-turbo, GPT-3.5-turbo-0301, and GPT-3.5-turbo-0613.
Among them, GPT-3.5-turbo is the latest model. On the other hand, GPT-3.5-turbo-0301 and GPT-3.5-turbo-0613 are deprecated snapshots from March 1st and June 13th of 2023, respectively.

\begin{table}[t]
\centering
\caption{Effectiveness and Efficiency when the maximum number of segments vary} \label{tab:K}
\vspace{-0.3cm}
\resizebox{0.8\columnwidth}{!}{%
\begin{tabular}{@{}lccccc@{}}
\toprule
\textbf{Max. Number of Segments} &  \textbf{1} &  \textbf{5} &  \textbf{10} & \textbf{15}   & \textbf{20}\\ \midrule
 Validation EM (\%) &12.9 &25.3    &27.8 &27.8  &27.6 \\
 Testing EM (\%) &10.2  &19.8     &20.0  &20.1  &19.9 \\
Inference Time per Function (s) &0.7 & 1.2  &1.9   &2.8 &4.1   \\ \bottomrule
\end{tabular}
}
\end{table}

\vspace{0.1cm}
\noindent\textbf{Results of Varying Maximum Number of Segments.}
Table~\ref{tab:K} presents the experimental results.
In previous experiments, we chose $K=10$ as it adequately covered over 90\% of the data samples.
From Table~\ref{tab:K}, we further validate that \textit{\textbf{K=10 strikes a balance between effectiveness and efficiency.}}
On one hand,  when $K$ is less than 10, VulMaster's effectiveness drops accordingly. For example, with $K=5$, the validation and testing EM scores decrease by 9.9\% and 1.0\%, respectively.
On the other hand, when $K$ is larger than 10, there is no further improvement in VulMaster's test EM scores. 
This is likely because $K=10$ already covers all input segments of over 90\% of data samples, and increasing it further does not provide significant benefits. 
However, higher $K$ leads to slower inference.
$K=10$ reaches a balance between effectiveness and efficiency.

\vspace{0.1cm}
\noindent\textbf{Results of Varying Prompts and ChatGPTs.}
The experimental results are shown in the bottom part of Table~\ref{tab:prompts}.
For three different prompts and different ChatGPT, we observe that VulMaster performance varies from 19.8\% to 20.3\%, with at most 3\% differences.
This indicates that \textit{\textbf{VulMaster can deliver satisfactory performance with varied prompts and ChatGPT models.}}

\vspace{0.2cm}
\noindent
\begin{tcolorbox} [boxrule=0.8pt,
                top=0.2pt,
                  bottom=0.2pt]
    \textbf{Answer to RQ3}:
    By setting the maximum number of segments as 10, VulMaster achieves a balance between effectiveness and efficiency. Besides, VulMaster shows stable performance across multiple plausible prompts and ChatGPT model versions, with at most 3\% differences.
\end{tcolorbox}

%% file: Sections/6_discussion.tex
\section{Discussion}
\label{sec:discussion}

\begin{table}[t]
\centering
\caption{Different prompts and the performance of VulMaster when prompts and ChatGPT models vary} \label{tab:prompts}
\vspace{-0.3cm}
\resizebox{0.95\columnwidth}{!}{%
\begin{tabular}{@{}llccc@{}}
\toprule
\multirow{12}{*}{\begin{tabular}[c]{@{}l@{}}\textbf{Prompt}\\ \textbf{Details}\end{tabular}} &  & \multicolumn{3}{l}{\textbf{Multiple Prompts}} \\ \cmidrule(l){2-5} 
 & \cellcolor[HTML]{EFEFEF} P0 & \multicolumn{3}{l}{\cellcolor[HTML]{EFEFEF}\begin{tabular}[c]{@{}l@{}}The code \{code\} contains a vulnerability of type \{name\}. \\ The analysis of this vulnerable code is \{analysis\}. \\ Please generate the repaired code to address the vulnerability:\end{tabular}} \\ \cmidrule(l){2-5} 
& \cellcolor[HTML]{ECF4FF} P1 & \multicolumn{3}{l}{\cellcolor[HTML]{ECF4FF}\begin{tabular}[c]{@{}l@{}}The code \{code\} suffers from \{name\}. \\ Here is the analysis for the code: \{analysis\}.\\ Please output the repaired code to fix the vulnerability:\end{tabular}} \\ \cmidrule(l){2-5} 
 &\cellcolor[HTML]{FFFBFB} P2 & \multicolumn{3}{l}{\cellcolor[HTML]{FFFBFB}\begin{tabular}[c]{@{}l@{}} \{code\} contains \{name\}. \{analysis\}. \\ Please generate the repaired code to address the vulnerability:\end{tabular}} \\ \bottomrule
\multirow{5}{*}{\textbf{Results}} & EM & gpt-3.5-turbo-0301 & gpt-3.5-turbo-0613 & gpt-3.5-turbo (latest) \\ \cmidrule(l){2-5} 
 & P0 &19.9  &20.1  &20.0  \\
 & P1 &20.0 &20.1  &20.1  \\
 & P2 &19.8  &20.3  &20.1  \\ \bottomrule
\end{tabular}
}
\vspace{-0.2cm}
\end{table}

\begin{figure}[b] 
\centering 
\vspace{-0.4cm}
\includegraphics[width=1\linewidth]{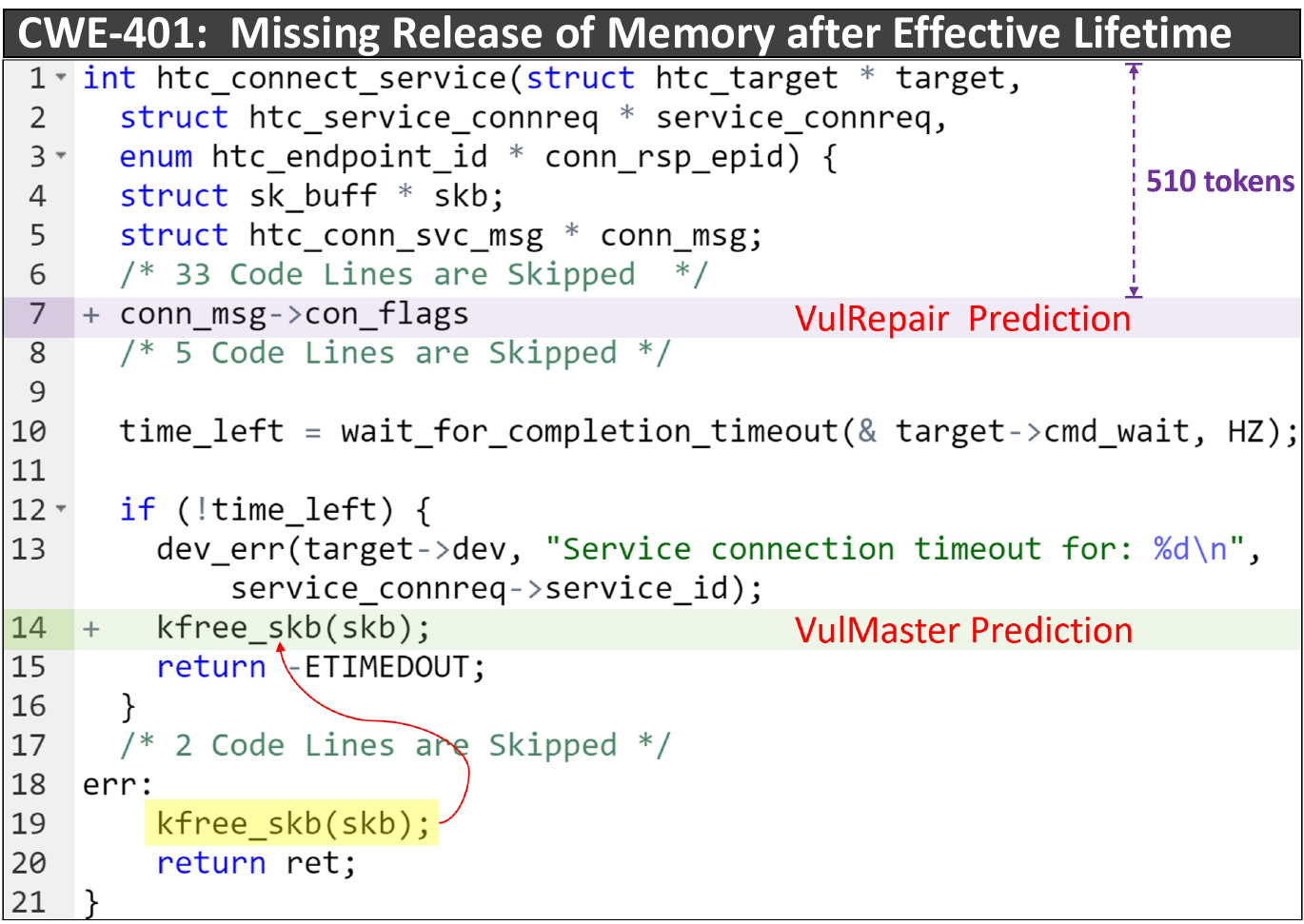} 
\vspace{-0.7cm}
\caption{Example of repairs by VulMaster and VulRepair.}
\label{fig:case} 
\end{figure}

\subsection{Case Study}

Figure~\ref{fig:case} presents two fixes generated by the SOTA VulRepair and VulMaster for a vulnerable function~\cite{qualitiaive_example} of CWE-401 from the evaluation dataset. CWE-401 refers to situations where memory is allocated dynamically but not properly released after it is no longer needed, leading to memory leaks~\cite{cwe401}.
The vulnerable function waits for the completion of a prior operation (\textit{Line 10}), and if the operation times out (\textit{Line 12}), it returns with an error code (\textit{Line 15}). However, it fails to release the dynamically allocated buffer ({\tt skb}) before the return statement, resulting in a CWE-401 vulnerability.

VulMaster correctly generates the fix (\textit{Line 14}) to free the memory with {\tt skb}. Its understanding of the entire function enables it to recognize the error-handling mechanism already present in \textit{Line 19} (i.e., {\tt kfree\textunderscore skb(skb);}), helping it to choose the suitable function (i.e., {\tt kfree\textunderscore skb}) for freeing the memory.
In contrast, the SOTA VulRepair makes a wrong prediction because the vulnerable function exceeds VulRepair's input limit of 512 code tokens before the vulnerability location (\textit{Line 14}). As a result, VulRepair's prediction only adds an irrelevant code line (\textit{Line 7}).

\begin{table*}[t]
\caption{The numbers of correctly fixed vulnerabilities by LLMs, APR models, and our VulMaster in the Vul4J~\cite{vul4j} benchmark 
} 
\label{tab:vul4j_result}
\vspace{-0.4cm}
\begin{center}
\footnotesize
\resizebox{2\columnwidth}{!}{%
\begin{tabular}{lr@{\quad}r@{\quad}r@{\quad}r@{\quad}r@{\quad}r@{\quad}r@{\quad}r@{\quad}r@{\quad}r@{\quad}r@{\quad}r@{\quad}r@{\quad}c}
\toprule
& \multicolumn{5}{c}{\textbf{Frozen LLMs}} & \multicolumn{4}{c}{\textbf{Fine-tuned LLMs}}  & \multicolumn{4}{c}{\textbf{APR models}} & \multicolumn{1}{c}{\textbf{Ours}}  \\
\cmidrule(lr){2-6}\cmidrule(lr){7-10}\cmidrule(lr){11-14}\cmidrule(lr){15-15}

    & Codex & CodeT5 & CodeGen & PLBART & InCoder & CodeT5 & CodeGen & PLBART & InCoder & CURE & Recoder & RewardR & KNOD & VulMaster 
    \\

\midrule
\textbf{Vul4J  (35)} & 6.2 & {2} & {1} & {0} &{3}& {2}& {5}& {2}& {6}& {1} & {0} & {0}  &   1 & \textbf{9} \\ 
\bottomrule
\end{tabular}
}
\end{center}
\vspace{-0.2cm}
\end{table*}

\subsection{How Do Fixes of CWE Examples Help?}

The ablation study (RQ2) revealed the effectiveness of the CWE knowledge.
The potential reason behind the effectiveness lies in that the generated fixes do not only describe ``what is vulnerable'' like the CWE names and vulnerable examples, but also explicitly provide knowledge on ``how to fix an identified vulnerability''. This is crucial information for the vulnerability repair task.
The validity of this explanation depends on the accuracy of the generated fixes.

To validate the explanation, the first and second authors manually assessed the correctness of the generated fixes for C/C++ vulnerable code examples. They carefully examined the vulnerable examples and the expert analyses from the CWE website and then reviewed the potential fixes generated by ChatGPT. Fixes that successfully mitigated the vulnerability were labeled as ``correct,'' while those that did not were labeled as ``wrong.'' In case of any discrepancies in annotations, a separate meeting was conducted to resolve each discrepancy, resulting in the final labels (shared in our replication package~\cite{vulmaster_rep}).
The manual investigation revealed that ChatGPT achieved an accuracy of \textbf{76.1\% (67/88)} in generating fixes for C/C++ vulnerable examples from CWE. This high accuracy supports our explanation to some extent.

Additionally, ChatGPT performs exceptionally well in generating fixes for vulnerable examples from CWE. However, it struggles to repair vulnerabilities from real-world projects (RQ1). The reason is that security experts have already provided ample analyses and descriptions of the vulnerable examples on the CWE website (as depicted in \ding{182} of Figure~\ref{fig:cwe_extraction}). 
In contrast, such detailed information is often unavailable when repairing vulnerabilities in real-world projects, making the task much more challenging.

\subsection{Generalizability to Benchmark with Tests}

\noindent
\textbf{Motivation.}
In our previous experiments, we utilized evaluation data from CVEFixes~\cite{cvefixes} and Big-Vul~\cite{fan2020ac}, encompassing a wide range of vulnerabilities collected from 1,754 C/C++ projects. However, this evaluation data solely included vulnerable code snippets and their associated fixes, lacking accompanying test cases. To ensure VulMaster’s generalizability to benchmarks that include test cases, we conducted an additional evaluation on the Vul4J~\cite{vul4j} benchmark. Vul4J~\cite{vul4j} is a high-quality Java vulnerability repair benchmark and includes test cases. Our evaluation of Vul4J also demonstrates VulMaster’s versatility in handling multiple programming languages, such as C/C++~\cite{cvefixes,fan2020ac} and Java~\cite{vul4j}.

\vspace{0.07cm}
\noindent
\textbf{Setup.}
The Vul4J benchmark serves as a test dataset for evaluation and does not provide a training pipeline for learning-based approaches. 
To overcome this limitation, we adopted the methodology of a recent study by Wu et al.~\cite{wu2023ective}, which conducted experiments comparing multiple Large Language Models (LLMs) and Automatic Patch Repair (APR) models on the Vul4J benchmark. 
Wu et al.~\cite{wu2023ective} conducted their evaluation by specifically selecting 35 single-hunk vulnerabilities from the Vul4J benchmark, as the APR models under investigation were primarily designed to address single-hunk bugs. Their comparative analysis involved three main categories: 1) the frozen LLMs, 2) fine-tuned LLMs, and 3) state-of-the-art APRs. The frozen LLMs directly performed inferences on the test set. For the fine-tuned LLMs, Wu et al. addressed the limited availability of Java vulnerabilities for fine-tuning by utilizing the general bug-fixing Java data shared in~\cite{Jiang2023ImpactOC} as their fine-tuning dataset. 
We utilized the bug-fixing Java data~\cite{Jiang2023ImpactOC} as the training data and followed Wu et al. to examine the correctness of the generated patches by using the available test cases and manual verification.

\vspace{0.07cm}
\noindent
\textbf{Results.}
Considering that the recent work~\cite{wu2023ective} utilized large model sizes for their baselines (up to 12 billion parameters), 
we introduced a larger variant of VulMaster in this discussion to align with the baselines. We use CodeT5p-large (0.8 billion parameters) as the backbone model in this experiment.
The results, presented in Table~\ref{tab:vul4j_result}, clearly demonstrate that VulMaster achieved the highest performance. Out of 35 vulnerabilities, VulMaster repaired 9, achieving a fix rate of 25.7\%.

\subsection{Explanations for Low Fix Rate}
VulMaster achieves a relatively low fix rate, primarily due to the inherent complexity of certain vulnerabilities. To delve into the causes of these incorrect fixes, we conducted a manual analysis. We randomly sampled 100 vulnerable-fixed code pairs from the test set and examined the corresponding generated fixes. 
Out of the 100 sampled vulnerabilities, VulMaster produced 78 incorrect fixes. Our manual investigation of these 78 failures revealed major causes as follows: 
1) Vulnerabilities requiring multiple-hunk code changes pose a significant challenge for VulMaster. This category accounted for 31 failures. Generating accurate multi-hunk patches remains a challenging task for learning-based models. Notably, many state-of-the-art APR techniques~\cite{jiang2021cure,zhu2021syntax,ye2022neural,knod} do not support bugs that require multi-hunk code changes. 
2) For 13 failures, developers’ correct patches incorporate members of structure/union variables that are not defined within the vulnerable functions, which means those members are unknown to VulMaster. Thus, it is unlikely for VulMaster to generate the correct fixes. 
3) For 8 failures, VulMaster fails to produce identical strings as the developers’ correct patches. 
4) For 5 failures, VulMaster fails to generate complex conditions in the correct fixes. 
5) For 5 failures, VulMaster fails to use the user-defined APIs correctly. 
6) For 4 failures, VulMaster uses the wrong APIs. 
7) For 5 failures, VulMaster generates semantically equivalent but syntactically different patches. Since we use Exact Match as the evaluation metric, these patches are considered ‘incorrect’ fixes.
Besides, 7 other failures cannot be classified.

\subsection{Threats to Validity}

Threats to internal validity pertain to potential errors and biases in our experiments. To address these threats, we take several precautions. Firstly, we strictly follow the settings and methodologies employed by baselines, using their official implementations to ensure consistency. We have reviewed and validated our code and data, which are publicly accessible for transparency and reproducibility.
Threats to external validity relate to the generalizability of VulMaster. To mitigate this potential concern, we meticulously select the experimental dataset, metrics, and baselines. The vulnerability repair dataset comprises 1,754 open-source software projects and has been utilized in prior learning-based state-of-the-art approaches. 
For the metrics, we select the EM, BLEU, and CodeBLEU, which are widely used in generation-based SE studies, e.g., ~\cite{DBLP:conf/icse/LiLLJHH23,DBLP:conf/icse/MahbubSR23}.
Lastly, we follow prior works~\cite{vrepair,vulrepair,wu2023ective} and assume perfect fault localization. We acknowledged that obtaining perfect or near-perfect localization is challenging. We consider more effective fault localization to be beyond the scope of this paper. We consider a human-in-the-loop setting where an experienced software engineer sifts through the output of a fault localization tool and only employs our approach once the location of the vulnerability is identified. We encourage future work to continue looking into better methods to localize faults, extending the popular line of work on fault localization.

%% file: Sections/7_related.tex
\section{Related Work}

\vspace{0.1cm}
\noindent\textbf{Vulnerability Datasets.}
Previous research has introduced various datasets to facilitate the evaluation of vulnerability repair approaches. The ManyBugs~\cite{le2015manybugs} dataset consists of 185 defects in 9 open-source programs. Most of their defects are logical errors rather than security vulnerabilities. 
The VulnLoc~\cite{vulnloc} dataset contains 43 vulnerable programs from 10 projects that span 6 CWEs. The Vul4J~\cite{vul4j} dataset includes reproducible vulnerabilities from 51 open-source Java projects, representing 25 different CWEs. CVEFixes~\cite{cvefixes} and Big-Vul~\cite{fan2020ac} are two large vulnerability repair datasets without test cases, curated from 1,754 projects. The large number of projects in  CVEFixes and Big-Vul datasets help ensure the diversity of vulnerabilities stored in them.

\vspace{0.1cm}
\noindent\textbf{Learning-based Vulnerability Repair.}
A number of learning-based approaches for repairing vulnerabilities have been proposed by researchers, extending the line of work on learning-based program repair, e.g.~\cite{le2016history}. 
For example, Ma et al.~\cite{DBLP:conf/esorics/MaTLSD17} proposed Vurle, 
the first learning-based approach for vulnerability repair; it learns transformative edits and their contexts from examples of vulnerable code fragments and their repairs.
Chi et al.~\cite{chi2022seqtrans} introduced SeqTrans, a Transformer-based machine translation model with copy mechanisms designed to fix Java vulnerabilities. Chen et al.~\cite{vrepair} proposed VRepair, which initially pre-trains a vanilla Transformer model on a bug-fixing corpus and subsequently utilizes it to address C/C++ vulnerabilities.
Fu et al.~\cite{vulrepair} proposed VulRepair, which leverages a BPE tokenizer and CodeT5 that is pre-trained on a large code corpus.
Wu et al.~\cite{wu2023ective} conducted an evaluation of nine large language models and four program repair models on the Vul4J dataset~\cite{vul4j}.
In contrast, VulMaster is specifically designed to effectively utilize diverse input sources: 1) input vulnerable functions, 2) code structures, and 3) rich CWE expert knowledge, while leveraging the power of Large Language Models. 
Our goal is to advance the progress of learning-based vulnerability repair approaches.

\vspace{0.1cm}
\noindent\textbf{Program Analysis-based Vulnerability Repair.}
Several program analysis-based approaches for repairing vulnerabilities have been proposed. 
CDRep~\cite{DBLP:conf/ccs/MaLLD16}, one of the earliest program analysis-based vulnerability repair approaches, employs a set of templates and data flow analysis to fix cryptographic-related vulnerabilities in Android apps, with a success rate of 90\%.
SenX~\cite{senx} aims to repair vulnerabilities by leveraging vulnerability-specific and human-specified safety properties. SAVER~\cite{saver} and Memfix~\cite{memfix} are designed to address memory errors. 
FootPatch~\cite{van2018static} generates patches that adhere to specific heap properties specified using separation logic. FootPatch is constrained to addressing only a limited set of common vulnerability types, such as memory leaks. VulnFix~\cite{zhang2022program} utilizes counterexample-guided inductive inference for repairing vulnerabilities. However, VulnFix is not applicable to vulnerabilities that cannot be resolved by modifying or inserting conditions, or situations that necessitate the introduction of new program variables~\cite{zhang2022program}. Unlike CDRep, SenX, SAVER, Memfix, FootPatch, and VulnFix, VulMaster is not limited to specific vulnerability types. Fix2Fit~\cite{gao2019crash} uses fuzz testing to filter out patches (produced by the APR model) that cause crashes. Combining VulMaster and Fix2Fit has the potential to enhance overall performance. CPR~\cite{cpr} employs concolic execution, along with a user-provided specification, to generate new inputs that aid in identifying and filtering out overfitting patches among the generated ones. 
ExtractFix~\cite{extractfix} uses dependency analysis and symbolic executions to fix vulnerabilities. 
In contrast to CPR and ExtractFix, VulMaster does not depend on heavy concolic and symbolic executions, making it able to generate patches more efficiently.

VulMaster differentiates itself from program analysis-based vulnerability repair approaches in several key aspects. Firstly, as a learning-based approach, VulMaster is not limited to a specific type of vulnerability, distinguishing it from various program analysis-based methods like SenX~\cite{senx}, SAVER~\cite{saver}, Memfix~\cite{memfix}, FootPatch~\cite{van2018static}, and VulnFix~\cite{zhang2022program}. 
However, for vulnerability types where good examples are hard to find, and specialized algorithms, precise templates, or formal safety properties are available, program analysis-based solutions will perform better. Thus, the two lines of work (learning-based and program analysis-based vulnerability repair) are complementary.
Secondly, VulMaster is adaptable to multiple programming languages, different from many program analysis-based vulnerability repair methods (such as SAVER~\cite{saver}, Fix2Fit~\cite{gao2019crash}, and ExtractFix~\cite{extractfix}) focusing solely on C/C++. 
Thirdly, VulMaster can generate patches more efficiently (about 10-20 seconds for one vulnerability) than program analysis-based vulnerability repair methods relying on heavy symbolic and concolic executions (e.g., CPR~\cite{cpr} and ExtractFix~\cite{extractfix}) or time-consuming dynamic invariant inference (e.g., VulnFix~\cite{zhang2022program}).

%% file: Sections/8_conclusion.tex
\section{Conclusion and Future Work}
\label{sec:conclusion}

We propose VulMaster, a Transformer-based vulnerability repair model through data-centric innovation and LLM collaboration. 
In our data-centric approach, we introduce a methodology that utilizes and combines various types of input data, including complete vulnerable code of any size, vulnerable code structures, and expert knowledge from the CWE system.
Moreover, we establish collaborations between two distinct Large Language Models, CodeT5 and ChatGPT, aiming to harness the strengths of each. CodeT5 is heavily customized to effectively absorb the knowledge from the training data, while ChatGPT generates the missing data for CodeT5.
The experimental results demonstrate that VulMaster outperforms all baselines significantly.
A replication package is provided at \textbf{\url{https://github.com/soarsmu/VulMaster_}}.

In the future, our interests will extend across multiple directions. As there is potential redundancy or noise in the inputs, we plan to design a refinement and denoising module to mitigate these issues.
We also plan to investigate additional evaluation metrics, c.f.,~\cite{DBLP:journals/jss/EvtikhievBSB23,DBLP:conf/wmt/Popovic15}. 
Moreover, we are keen on employing larger pre-trained code models for this task in the future, utilizing parameter-efficient tuning techniques like~\cite{weyssow2023exploring}. 
Additionally, we plan to develop {\em trustworthy and synergistic} AI4SE methodologies~\cite{DBLP:journals/corr/abs-2309-04142} to help software engineers better trust and synergize with VulMaster and other automated vulnerability identification and repair solutions.

\vspace{0.2cm}
\noindent \textbf{Acknowledgement.}  This research / project is supported by the National Research Foundation, under its Investigatorship Grant (NRF-NRFI08-2022-0002). Any opinions, findings and conclusions or recommendations expressed in this material are those of the author(s) and do not reflect the views of National Research Foundation, Singapore.